\documentclass[proceedings]{ccn}
   \ccndoi{10.32470/06d965e7}  

\renewcommand\vec{\boldsymbol}

\title{Implications of hierarchical Markov models of behavior: \\on irreversibility, predictability, and dimensionality}

 \author{John J. Vastola and Kanaka Rajan \\
   Department of Neurobiology, Harvard Medical School \\Kempner Institute for the Study of Natural and Artificial Intelligence, Harvard University \\
   \email{\{john\_vastola, kanaka\_rajan\}@hms.harvard.edu}}

\begin{document}

\maketitle

  
\begin{abstract}
The maturation of quantitative tools for studying the high-level structure of animal behavior, and especially tools which represent spontaneous behavior as a sequence of stereotyped and neurally well-defined `syllables', demands that the field revisit a fundamental theoretical question: if the coarse structure of behavior can be accurately described by Markov models, what do these models really tell us about behavior? In this work, we explore the theoretical implications of these models and discuss how they allow us to quantitatively formulate questions about the sequence-like nature and effective dimensionality of behavior. One important insight is that the eigenvalues and eigenvectors of various model-associated matrices furnish interpretable time scales and modifications of behavior that occur on those time scales. We illustrate our points using both toy examples and Markov models fit to real data. By analyzing the consequences of Markov representations, we clarify the theoretical meaning of progress in quantifying behavior.
\end{abstract}

\section{Introduction}

Understanding the high-level structure of animal behavior---particularly the spontaneous and innate behavior associated with naturalistic environments \citep{tinbergen_study_1951}---has only recently become quantitatively tractable. Techniques for the large-scale quantification of behavior, including algorithms for tracking animal positions and poses for at least tens of hours \citep{mathis_deeplabcut_2018,pereira_sleap_2022,pereira_quantifying_2020}, enable us to precisely formulate and answer questions about how behavior evolves across contexts. One revelation from these quantitative ethology studies is that behavior is perhaps simpler than anyone might have expected: there is good evidence that behavior is not infinitely variable, but consists of fairly stereotyped sequences of discrete \textit{syllables} \citep{wiltschko_mapping_2015,weinreb_keypoint-moseq_2024}. Simple syllables, like `walking' or `sniffing', can be chained together to form more complex behaviors, like investigating objects or exploring an environment.

In the syllable-based view, behavior is quantitatively characterized via Markov chain models with finitely many states. Animals behave \textit{as if} there is a constant probability per unit time of transitioning to a different discrete behavior, in a fashion generally modulated by both the current behavior and some kind of latent state (e.g., hunger). Older efforts to construct Markov models of behavior emphasized the poor quality of their fit to data (or equivalently, signatures of `non-Markovianity'), which was interpreted as a sign that unobservable latent states acted as confounds \citep{schwarz_changes_2015,berman_fly_2016,overman_measuring_2022}. But more recent efforts have shown that Markov models provide an accurate description of the behavior of \textit{C. elegans} \citep{costa_elegans_2024}, larval zebrafish \citep{gautam_zebrafish_2024}, and mice \citep{weinreb_spontaneous_2026}, provided one explicitly accounts for latent states. The work of \citet{weinreb_spontaneous_2026} is particularly noteworthy, because in addition to identifying behavioral syllables and latent states which modulate them, it shows that these syllables and latent states have neural correlates in the dorsomedial prefrontal cortex. When considered in light of the earlier finding that the dorsolateral striatum encodes behavioral syllables \citep{markowitz_spontaneous_2023}, one finds strong support for the claim that Markov models capture biologically meaningful structure rather than arbitrary abstractions.

It stands to reason that these techniques will improve and lead to even better Markov models of behavior. These improvements highlight a \textit{theoretical} question more subtle than questions concerning fit quality: what do these models \textit{actually tell us} about the high-level structure of behavior? In this work, we argue that drawing upon the rich mathematical literature associated with Markov models yields interesting tentative answers to fundamental questions about behavior. Here, we focus especially on questions regarding (i) the sequence-like character of behavior, and (ii) the effective dimensionality of behavior, but our point is in principle broader. A key insight is that spectral decompositions associated with Markov chains can provide interpretable time scales and modes of behavior, which are `global' in the sense that they depend on the high-level structure of the Markov chain rather than on just a small number of syllable transition rates. But there is not just one informative decomposition---there are at least two, and each is naturally associated with a somewhat different question. 

We will begin by reviewing the mathematics of Markov models and their application to behavior. In the following sections, we consider the sequence-like character of behavior, the dimensionality of behavior, and the consequences of hierarchical structure. Throughout, we illustrate our points using both carefully chosen toy models and a model fit to data from \citet{weinreb_spontaneous_2026}. 

\section{Markov models of behavior}

\begin{figure*}[ht]
  \begin{center}
   \includegraphics[width=\textwidth]{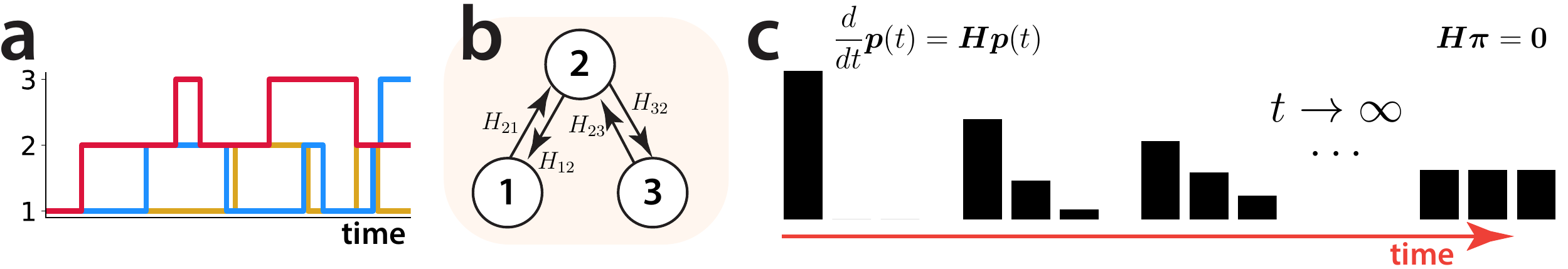}
  \end{center}
  \caption{\textbf{Markov chain models of behavior.} \textbf{(a)} Transitions between different discrete behaviors (here, labeled 1, 2, and 3) are not random, but structured. Different colors represent different stochastic simulations. \textbf{(b)} Markov models of behavior can be visualized as graphs whose nodes represent behaviors and behavioral states, and whose arrows reflect transition rates. In the pictured 3-state model, it is impossible to transition directly between 1 and 3. \textbf{(c)} The transition matrix $\vec{H}$ defines a dynamics in probability space (here: the space of distributions of 3 behaviors). All Markov models we consider eventually converge to a unique steady state distribution $\vec{\pi}$ (which here is uniform).}
  \label{fig:schematic}
\end{figure*}

\subsection{The discreteness of behavior}

The primitive insight that suggests Markov models of behavior is that, although behavior certainly involves many approximately \textit{continuous} degrees of freedom---which implies that the number of possible behaviors is effectively infinite---there is on the other hand a great deal of stereotypy. This is especially apparent when behavior is suitably coarse-grained, so that slight variations of one movement (e.g., a reach) are treated as identical. Following work on MoSeq \citep{wiltschko_mapping_2015,weinreb_keypoint-moseq_2024}, we will call discrete units of behavior \textit{syllables}, but the reader should note that essentially the same concept has been used with other names, like `eigenworms' \citep{stephens_dimensionality_2008} and `bout types' \citep{marques_structure_2018,johnson_probabilistic_2020}. 

Motivated by this insight, in what follows we ignore continuous structure and treat behavior as discrete. We consider models with $N_B \geq 1$ possible behaviors, and use an integer $b \in \{ 1, ..., N_B \}$ to index each behavior.

\subsection{From bags to bigrams to Markov models}

The simplest discrete model of behavior is probably a `bag of behaviors' model akin to `bag of words' models of language \citep{wallach_2006_wordbag}. One can construct such a model by observing an animal for a long period of time $T$, and then determining the total amount of time $T_b$ it spent performing each behavior $b$. Assuming behavioral syllables are defined in a way that prevents the animal from doing more than one at a time, and that the set of $N_B$ behaviors covers all possibilities, the \textbf{bag model} defines the probability of performing $b$ as $\pi_b := T_b/T$. This model decides what to do next by `drawing from the bag' according to the probabilities $\vec{\pi} := (\pi_1, ..., \pi_{N_B})^T$, regardless of the current behavior.

Although this model has some use in naturalistic settings \citep{altmann_observational_1974}, particularly since it requires less data to fit than more sophisticated models (there are only $N_B-1$ free parameters), it is a bad model of behavior. It is bad principally because it neglects \textit{transition} structure (Fig. \ref{fig:schematic}a): for example, blowflies alternate between front leg grooming and anterior grooming movements in a stereotyped fashion \citep{richard_hierachical_1976}.

The most obvious way to improve the bag model is to instead compute \textbf{bigrams}: how often behavior $i$ follows behavior $j$. To measure these, divide the total amount of time $T$ an animal behaves into time bins, each of size $\sim \Delta t$. This `mesoscopic' time scale $\Delta t$ must be chosen small enough that a single behavior dominates each bin. One then obtains the transition probabilities $p_{ij}$ by counting the number of times an $i$ bin follows a $j$ bin.

\textbf{Markov models} of behavior assume bigram structure completely characterizes dynamics. We can construct a continuous-time model from bigram statistics by assuming, for small enough $\Delta t$, that the probability $p_{ij}$ of a $j$ to $i$ transition scales linearly with $\Delta t$, i.e., $p_{ij} = k_{ij} \Delta t + \mathcal{O}((\Delta t)^2) \approx k_{ij} \Delta t$. We can then construct 
\begin{equation}
H_{i j} := \begin{cases}
- \sum_{b \neq  i} k_{b \ i} & \text{ for } i = j \\
k_{ij} & \text{ for } i \neq j \ ,
\end{cases}
\end{equation}
which we call a \textbf{transition matrix}. $\vec{H}$ is \textit{infinitesimal stochastic} since (i) its off-diagonal entries, which represent transition rates, are nonnegative; and (ii) its columns sum to zero \citep{baez_book_2018}. 

This matrix defines a continuous-time Markov chain model of behavior (Fig. \ref{fig:schematic}b, c) via the \textbf{master equation}
\begin{equation} \label{eq:markov_model_def}
\frac{d}{dt} \vec{p}(t) = \vec{H} \vec{p}(t) \hspace{0.2in} \text{ i.e., } \hspace{0.1in} \frac{d}{dt} p_b(t) = \sum_{a=1}^{N_B} H_{b a} p_a(t) 
\end{equation}
where $p_b(t)$ is the probability that the animal is performing behavior $b \in \{1, ..., N_B \}$ at time $t$, and $\vec{p}(t) := ( p_1(t), ..., p_{N_B}(t) )^T$. Note that, because the columns of $\vec{H}$ sum to zero, probability is conserved, since $\vec{1}^T \vec{H} = \vec{0}^T$ implies $\frac{d}{dt} \vec{1}^T \vec{p}(t) = \vec{1}^T \vec{H} \vec{p}(t) = \vec{0}^T \vec{p}(t) = 0$. 

Parenthetically, we note that using a continuous-time (rather than discrete-time) formulation has a number of advantages for studying behavior. In continuous-time formulations, it is natural to think in terms of bin-size-independent rates and time scales, whereas in discrete-time formulations model properties depend on $\Delta t$.

\subsection{What about non-Markovianity?}

It has been pointed out numerous times that bigram statistics are generally insufficient for describing real animal behavior \citep{schwarz_changes_2015,berman_fly_2016}, which is equivalent to the statement that Markov models are a poor description. But as previously mentioned, various authors have shown that the situation changes if latent states are included in the model's notion of `state' \citep{costa_elegans_2024,gautam_zebrafish_2024,weinreb_spontaneous_2026}. In this work, we take the fairly uncontroversial view that non-Markovianity is merely a sign that the state space ought to be enlarged.

\subsection{Technical assumptions restricting models}

We do not allow all possible models of behavior (or equivalently, all possible infinitesimal stochastic matrices $\vec{H}$), because some are degenerate in one or more senses. In brief (see Appendix A for more discussion), we assume: (i) there are no states that it is impossible to enter; (ii) there are no states it is impossible to leave;
(iii) all states are connected by some directed path through the transition graph; and (iv) $\vec{H}$ is diagonalizable. These assumptions ensure that there exists a well-defined \textbf{steady state distribution} $\vec{\pi} := \lim_{t \to \infty} \vec{p}(t)$ which Eq. \ref{eq:markov_model_def} always asymptotically approaches (Fig. \ref{fig:schematic}c). Up to scaling, $\vec{\pi}$ is also the unique (right) eigenvector of $\vec{H}$ with eigenvalue zero. In some sense, using the bag model amounts to throwing away $\vec{H}$ and keeping only $\vec{\pi}$.

\section{Irreversibility and action sequences}
\label{sec:irrev}

\begin{figure*}[ht]
  \begin{center}
   \includegraphics[width=\textwidth]{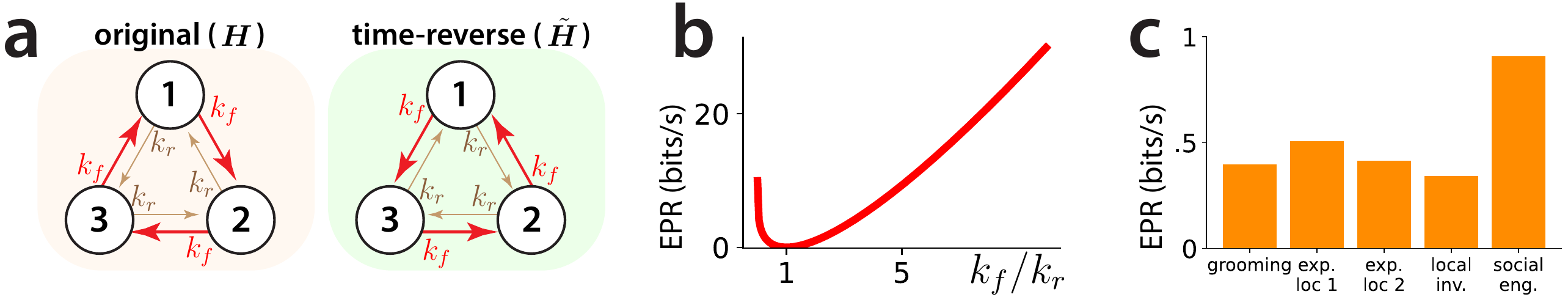}
  \end{center}
  \caption{\textbf{Quantifying irreversibility.} \textbf{(a)} One way to visualize the time-reverse of a Markov model is to reverse the arrows of the corresponding graph. However, note that $\tilde{H}_{ij} \neq H_{ji}$ in general. Pictured: the cyclic model with $N = 3$ states. \textbf{(b)} The entropy production rate (EPR) of the cyclic model equals $(k_f - k_r) \log_2(k_f/k_r)$. It increases as circulation around the ring becomes more directed. \textbf{(c)} The EPR differs for different latent states in a Markov model fit to open-field mouse behavior, and is highest in the `social engagement' state.}
  \label{fig:irrev}
\end{figure*}

A fundamental feature of animal behavior is that it involves sequences of actions with a fairly stereotyped ordering. For example, in fruit fly courtship a male typically orients toward a female, taps her abdomen with his forelegs, extends one wing, and then begins a courtship song \citep{villella_chapter_2008}. If we take Markov models seriously as a description of behavior, it is interesting to note that not \textit{all} possible models generate predictable sequences of actions; rather, different models do this more or less, in a way that can be quantified.

\subsection{Sequences reflect irreversibility}

One way of mathematically formalizing the idea that behavior consists of stereotyped sequences is via the concept of \textbf{irreversibility}: if we watch a movie of a behaving animal, it ought to look strange when played backwards. In the context of Markov models, we can define a precise notion of `playing the movie backwards' by defining the time-reverse of a behavioral Markov chain. If $k_{ij} \Delta t$ is the probability of moving from state $j$ to state $i$ in a small amount of time $\Delta t$, then by Bayes' rule the time-reversed chain has a small-time transition probability
\begin{equation}
\tilde{p}_{ij} := \frac{p_{j i} \pi_i}{\pi_j} = k_{j i} \frac{\pi_i}{\pi_j} \Delta t \ ,
\end{equation}
where we fix the prior by assuming the chain is stationary. Let $\vec{D}$ denote the diagonal matrix with $D_{ii} = \pi_i$ for all $i$. These probabilities can be associated with the transition matrix $\tilde{\vec{H}} := \vec{D} \vec{H}^T \vec{D}^{-1}$, called the \textit{adjoint} of $\vec{H}$, which is infinitesimal stochastic and hence defines its own Markov chain. This chain's graph looks just like the graph of the original chain, but with all arrows reversed (Fig. \ref{fig:irrev}a). As one might expect, for chains where transitions tend to be bidirectional (which one might call `diffusive'), the time-reversed chain tends to look similar to the original chain; meanwhile, chains which involve directed cycles tend to look quite different from their time-reverse.

\subsection{Detailed balance as a null hypothesis}

What is the \textit{least} sequence-like Markov model of behavior? If we accept that irreversibility is a reasonable way to quantify sequence-like-ness, then we can define a `null hypothesis' via the concept of \textbf{detailed balance}. Models satisfying detailed balance obey the condition
\begin{equation}
H_{ij} \pi_j = H_{ji} \pi_i
\end{equation}
for all pairs of behaviors $i$ and $j$, which is equivalent to asking that $\vec{H} = \tilde{\vec{H}}$ \citep{kelly2014stochastic}. This condition says that, when the chain is at steady state, any transition one observes is equally likely to have been observed in the opposite order. One cannot tell whether a movie of behavior was played `forwards' or `backwards', because there is no meaningful difference.

The concept of detailed balance will be useful to us for a variety of reasons, even beyond this discussion of irreversibility; we will see in the following sections that it makes it easier to define the dimensionality of behavior.

\subsection{How irreversible is behavior?}

We will now consider two questions. First, how can we measure the extent to which behavior is irreversible given a Markov chain description of it? Second, how irreversible is real animal behavior? Since irreversibility reflects the extent to which $\vec{H}$ differs from its adjoint $\tilde{\vec{H}}$, a simple way to quantify it is to measure the fluxes
\begin{equation}
J_{ij} = H_{ij} \pi_j - H_{ji} \pi_i = \pi_j ( H_{ij} - \tilde{H}_{ij} )
\end{equation}
for all pairs of behaviors $i$ and $j$. (Since $J_{ij} = - J_{ji}$, ordering does not matter.) In simple cases, we can compute $J_{ij}$ explicitly. For the \textbf{cyclic model}, which has $N$ states arranged along a ring so that `forward' transitions occur at a rate $k_f > 0$ and `reverse' transitions occur at a rate $k_r \geq 0$ (see Appendix B), $|J_{ij}|$ equals $|k_f - k_r|/N$ for neighboring states and zero otherwise. But while measuring fluxes has the advantage that it localizes irreversibility to specific sets of transitions, there are at least two problems: the $J_{ij}$ can be small because they scale with $N$, and it is not obvious how to compare them across systems. Alternatively, we can define an aggregate measure of irreversibility, with the \textbf{entropy production rate} (EPR) \citep{seifert_entprod_2005,lynn_db_2021}
\begin{equation}
\dot{S} :=  \sum_{i \neq j} H_{ij}  \pi_j \log_2\left(  \frac{H_{ij} \pi_j}{H_{ji} \pi_i}  \right)  
\end{equation}
a canonical choice. For chains satisfying detailed balance, all fluxes $J_{ij}$ are zero and the EPR is exactly zero. For the cyclic model $\dot{S} = (k_f - k_r) \log_2(k_f/k_r)$ (Fig. \ref{fig:irrev}b), so it increases as $k_f$ and $k_r$ are made increasingly different.

To briefly address the irreversibility of real behavior, we quantified the EPR associated with a hierarchical Markov model fit to open-field mouse behavior from \citet{weinreb_spontaneous_2026} (see Appendix C). This model has five high-level latent states which modulate behavior, labeled `grooming', `exploratory locomotion 1', `exploratory locomotion 2', `local investigation', and `social engagement'. We find that the $\vec{H}$ associated with each latent has a nonzero EPR, with the `social engagement' state having around double the EPR of the other states (Fig. \ref{fig:irrev}c). This suggests the unsurprising conclusion that real behavior is indeed at least somewhat irreversible, and can be more or less irreversible in different contexts.

\section{Dynamics and dimensionality}
\label{sec:dyndim}

\begin{figure*}[ht]
  \begin{center}
   \includegraphics[width=0.8\textwidth]{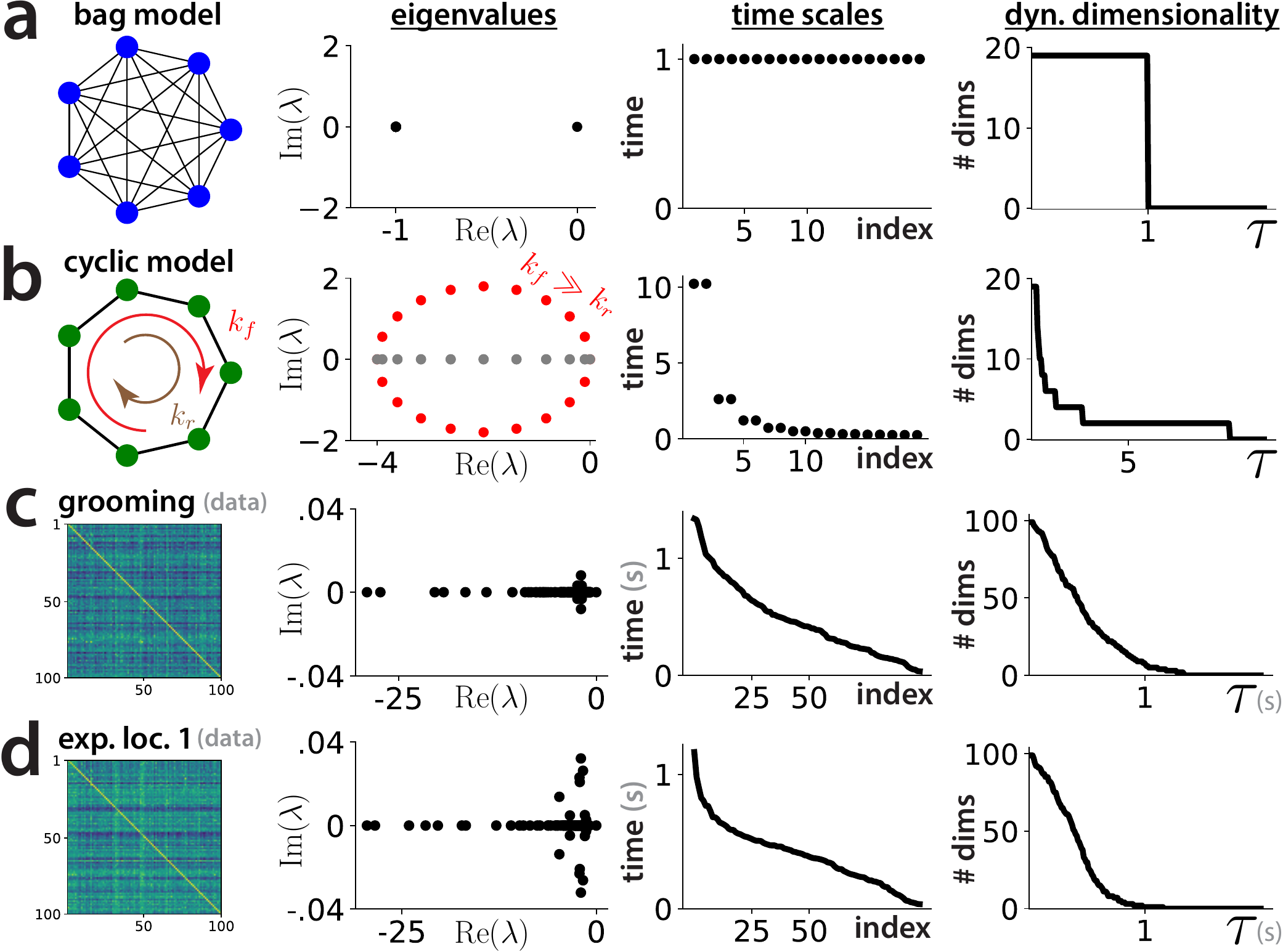}
  \end{center}
  \caption{\textbf{Dimensionality via dynamics.} From left to right: model schematic, eigenvalues $\lambda_k$ of $\vec{H}$, time scales $\tau_k := - 1/\text{Re}(\lambda_k)$ ($\lambda_k \neq 0$) ordered from greatest to least, and effective (dynamical) dimensionality as a function of the time scale $\tau$. Each row depicts these for a different model. \textbf{(a)} The bag model ($N = 20$ used) has just two distinct eigenvalues and one time scale, and its dimensionality is either maximal or zero. \textbf{(b)} The cyclic model ($N = 20$ used) has many distinct eigenvalues, which are real if $k_f = k_r$ (gray) but can be complex-valued if $k_f \neq k_r$ (red). The top two nonzero modes provide a reasonable approximation to the system for many choices of $\tau$, so the system is roughly two-dimensional. \textbf{(c)} Mice in a `grooming' state exhibit behavior involving a range of sub-second time scales, and dimensionality is not well summarized by a small number of modes. \textbf{(d)} In the `exploratory locomotion 1' state, the imaginary parts of some eigenvalues are larger, which suggests behavior is more irreversible.}
  \label{fig:dyndim}
\end{figure*}

Another fundamental feature of behavior is that it tends to be somewhat low-dimensional, in the sense that the actions and action sequences actually executed by animals occupy an extremely restricted region of the space of all possible actions and action sequences. Does a Markov model description of behavior tell us anything about this? While it cannot tell us about dimensionality in terms of the \textit{number} of actions animals can access, since by construction it only involves the actions an animal actually performed, it can say something interesting about the dimensionality of \textit{transitions} between behaviors, and hence provide a nontrivial notion of the dimensionality of behavior. We elaborate on this idea below.

\subsection{Intuition: distribution dimensionality}

Perhaps the most naive way to define the dimensionality of behavior is to consider grouping or coarse-graining behaviors themselves; in our Markov model picture, this corresponds to grouping syllables with similar connectivity, for example via graph clustering algorithms. On the other hand, in moving from continuous behavior to discrete syllables, some such coarse-graining (e.g., treating many different locomotion patterns as ``walk'' or ``run'') has already been done. It is not clear what useful information additional syllable-level coarse-graining tells us.

A different and arguably more useful notion of dimensionality concerns the behavior distribution $\vec{p}(t)$. Because $\vec{p}(t)$ is a (normalized) vector in $\mathbb{R}^{N_B}$, how $\vec{p}(t)$ evolves in time can be visualized as a trajectory in $\mathbb{R}^{N_B-1}$. Any change in context---for example, some change in the environment, or change in the animal's latent state---will cause $\vec{p}(t)$ to be different than $\vec{\pi}$, so that the Markov chain is out of equilibrium. As time passes, it will approach $\vec{\pi}$ along a path which depends on its initial condition. One way to pose the question of dimensionality is: for many different initial conditions, what is the dimensionality of these trajectories?

Since $\vec{p}(0)$ can in principle be anything, one may suspect the answer is always $N_B - 1$, the dimensionality of the whole space. But this is \textit{not} true if we \textit{additionally} coarse-grain over time, and only answer our question above a time scale $\tau \geq 0$. On long enough time scales, the space of trajectories generally looks lower-dimensional---and in the $\tau \to \infty$ limit, the space collapses to the point at $\vec{\pi}$. One way to visualize this is to imagine cutting out the beginning of many movies of behavior (i.e., frames before $\tau$), and then considering only the remaining frames of the trajectories.

For some moderate value of $\tau$, when do we expect the dimensionality to be lower than maximal? This can happen when there is a separation of time scales in behavioral transitions, which can cause some adjustments to the behavior distribution to happen quickly (or at least, faster than $\tau$) and others to happen more slowly. A 3-state model of sleep provides a trivial albeit informative example: animals may transition between `more awake' and `less awake' states relatively quickly (see Appendix B), but between these and a `sleep' state slowly. Beyond the fast time scale, the model `looks' like it has only two states (`awake' and `sleep'), and $\vec{p}(t)$ trajectories are effectively one-dimensional.

\subsection{Making this intuition rigorous}

We can formalize this intuition by invoking a mathematical result known as the Perron-Frobenius theorem. This theorem says that given certain technical assumptions that rule out degenerate cases (see Appendix A), the behavior distribution $\vec{p}(t)$ has a spectral decomposition
\begin{equation} \label{eq:spectral_decomp}
\vec{p}(t) = \vec{\pi} - \sum_{k=1}^{N_B - 1} c_k \vec{v}_k e^{\lambda_k t} = \vec{\pi} - \sum_{k=1}^{N_B - 1} c_k \vec{v}_k e^{- t/\tau_k + i \omega_k t} 
\end{equation}
where $\lambda_k = -1/\tau_k + \omega_k i$ is the $k$-th eigenvalue of $\vec{H}$, $\vec{v}_k$ is the $k$-th right eigenvector of $\vec{H}$, and the $c_k$ are initial-condition-dependent coefficients. This bears on trajectories through distribution space because the $\vec{v}_k$ form a basis for that space, and $\vec{v}_k$ only meaningfully contributes to $\vec{p}(t)$ on time scales shorter than $\tau_k$. Hence, we will say that the \textbf{dynamical dimensionality} of a model is $r$ on a time scale $\tau$ if there are $0 \leq r \leq N_B - 1$ eigenvectors whose time scale is at least $\tau$.

\subsection{Behavioral modes and time scales}

This result also provides an interesting formalization of behavioral modes and time scales. The components of each $\vec{v}_k$ sum to zero, so each eigenvector represents adjustments to the behavior distribution (i.e., do some behaviors more and others less) which happen on a time scale $\tau_k$. Consider, for example, the 3-state sleep-wake Markov chain: one eigenvector corresponds to equilibration of the awake states, and the other corresponds to movement between the awake and asleep states (see Appendix B). One interesting feature of these summaries is that they are `global': they do not depend only on specific behavioral transitions, but on the structure of transitions writ large. Generally, multiple transition rates $k_{ij}$ factor into any eigenvalue or eigenvector, as one can see by studying simple examples.

A Markov chain version of the \textbf{bag model} (see Appendix B) has maximal dimensionality until $\tau$ exceeds a critical value (Fig. \ref{fig:dyndim}a). The cyclic model, whose transitions are much more structured, is effectively two-dimensional for a broad range of $\tau$ (Fig. \ref{fig:dyndim}b). The model of open-field mouse behavior from \citet{weinreb_spontaneous_2026} involves a wider range of sub-second time scales than these toy models, and is not obviously low-dimensional unless $\tau$ is at least a second (Fig. \ref{fig:dyndim}c, d).

\section{Predictability and dimensionality}
\label{sec:preddim}

\begin{figure*}[ht]
  \begin{center}
    \includegraphics[width=0.8\textwidth]{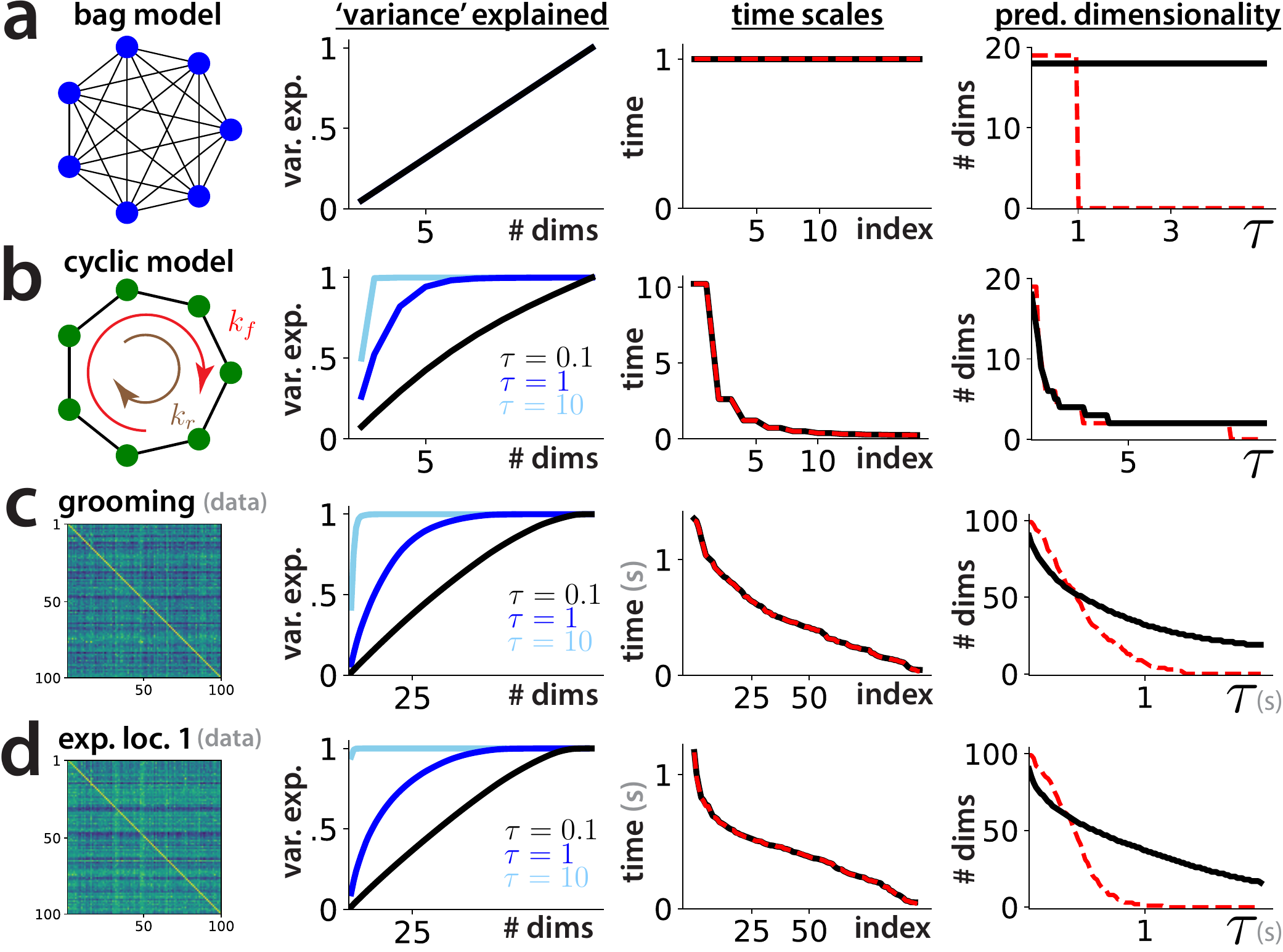}
  \end{center}
  \caption{\textbf{Dimensionality via predictability.} From left to right: model schematic, `variance' explained as a function of the number of modes used (different colors correspond to different values of the imposed time scale $\tau$), time scales $\tau_k := - 2 \tau/\log(\nu_k)$ ($\nu_k \neq 1$) ordered from greatest to least (red dashed line: time scales of $\vec{H}$), and predictive dimensionality vs $\tau$ (here, number of modes required to capture $\geq 90\%$ variance; red dashed line: dynamical dimensionality vs $\tau$). Each row depicts these for a different model. \ \textbf{(a)} Each mode of the bag model ($N = 20$ used) explains an equal amount of variance, and dimensionality is near-maximal. \textbf{(b)} A small number of modes explains most variance for the cyclic model ($N = 20$ used), especially for larger $\tau$. \textbf{(c-d)} Few modes explain most variance when $\tau$ is more than a few seconds, but predictive dimensionality tends to be larger than dynamical dimensionality. }
  \label{fig:preddim}
\end{figure*}

Unfortunately, the notion of dimensionality defined in the previous section is not without problems. It does not necessarily answer a related question: how many \textit{orthogonal} vectors are required to characterize changes in the behavior distribution beyond a time scale $\tau$? If detailed balance is violated and a model is highly irreversible, it is possible for its eigenvectors to be linearly independent but nearly aligned. This is essentially another manifestation of the well-known issue that non-normal dynamics are more complex than normal dynamics \citep{ganguli_normal_2008}. This issue motivates a different notion of dimensionality associated with predictability, which we call a model's \textbf{predictive dimensionality}.

\subsection{Predictability yields a PCA analogue}

Following Bialek \citep{bialek_predictability_2001,bialek_2022_behave}, a principled information-theoretic approach to our question requires that we quantify the mutual information between the past and future. But a key difficulty with this approach is that mutual information and associated quantities can be difficult to actually compute. An alternative is to approximate mutual information as a simpler quadratic form; one can show (see Appendix D) that the mutual information between the system's state at one time, and its state after $\tau$ additional time, is roughly
\begin{equation}
I(\tau) \approx \frac{1}{2} \text{tr} \vec{K}(\tau) - \frac{1}{2} \  ,
\end{equation}
where $\vec{K}(\tau) := \vec{D}^{1/2} e^{\vec{H}^T \tau} \vec{D}^{-1} e^{\vec{H} \tau} \vec{D}^{1/2}$ is the past-future kernel. Unlike $\vec{H}$, this matrix is always symmetric and positive definite, so its eigenvalues $\nu_k$ are all real and positive. This is important because $\vec{K}(\tau)$ can essentially play the role that the covariance matrix does in principal component analysis (PCA): we can approximate $\vec{p}(t)$ using eigenvectors of $\vec{K}(\tau)$ to capture as much `variance' as possible. Supposing eigenvalues are ordered from greatest to least, and that $\nu_k = 1$ is excluded, we say that the top $r$ eigenvectors explain the fraction
\begin{equation}
\text{ve}(\tau) := \frac{\nu_1 + \cdots \nu_r}{\sum_{k=1}^{N_B - 1} \nu_k} \leq 1
\end{equation}
of the overall variance. We say a model has \textbf{predictive dimensionality} $r$ on a time scale $\tau$ if $r$ is the minimum number of modes required for $\text{ve}(\tau)$ to exceed a desired threshold (e.g., $90\%$). Interestingly, predictive and dynamical dimensionality look similar for systems which obey detailed balance. Since $\vec{K}(\tau) = \vec{D}^{-1/2} e^{\tilde{\vec{H}} \tau} e^{\vec{H} \tau} \vec{D}^{1/2}$, detailed balance implies that $\nu_k = e^{2 \lambda_k \tau}$, and that the eigenvectors of $\vec{K}(\tau)$ are those of $\vec{H}$ up to a $\vec{D}^{-1/2}$ rescaling. Away from detailed balance, $\vec{K}(\tau)$ can be viewed as producing a certain `reversiblization' of the original model, which approximates it with a model that \textit{does} satisfy detailed balance \citep{choi_reversib_2024}.

\subsection{Behavioral modes and time scales}

The eigenvalues and eigenvectors of $\vec{K}(\tau)$ are all real, and hence easier to interpret: they are time scales on which a given orthogonal behavioral mode is relevant for predicting changes in the behavior distribution $\vec{p}(t)$, in the sense of maximizing predictability. Dynamical and predictive dimensionality look very similar for the toy models we consider (Fig. \ref{fig:preddim}a, b), but predictive dimensionality tends to be significantly larger than dynamical dimensionality for models fit to open-field mouse behavior (Fig. \ref{fig:preddim}c, d). We speculate that this is because, when $\vec{H}$ is highly irreversible and many of its eigenvectors are aligned, one must use more orthogonal vectors to capture the same amount of predictability.

\section{Consequences of hierarchy}
\label{sec:hierarchy}

\begin{figure*}[ht]
  \begin{center}
     \includegraphics[width=\textwidth]{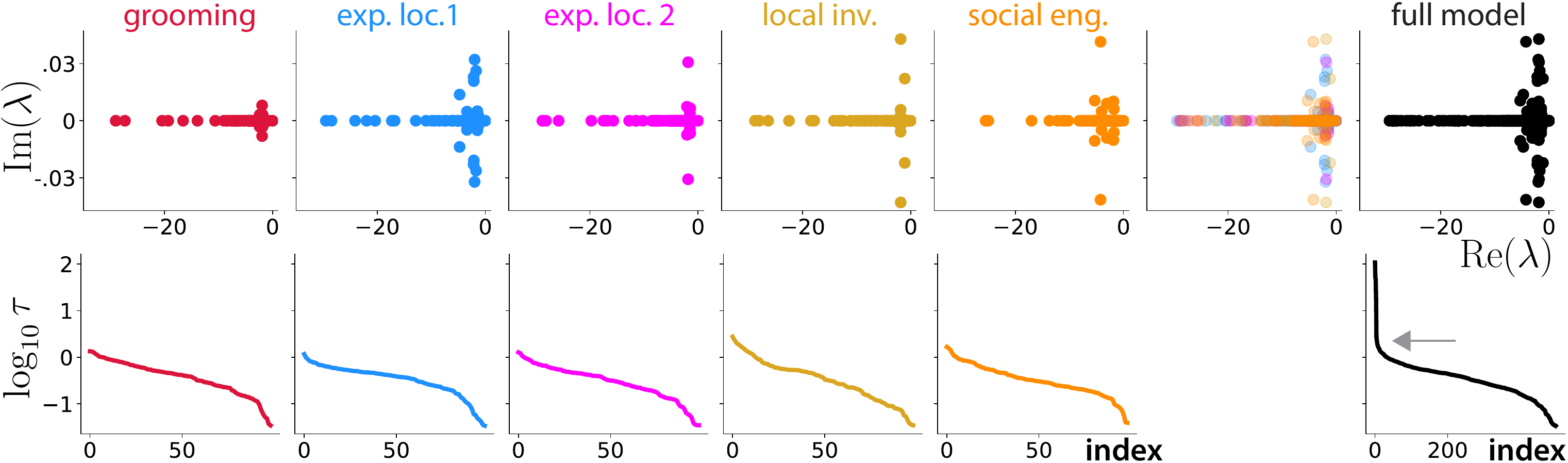}
  \end{center}
  \caption{\textbf{Hierarchical spectral structure in behavior.} The first five columns depict the eigenvalues of $\vec{H}$ (top row) and associated time scales (bottom row, ordered from greatest to least) in different latent states. The sixth column shows all eigenvalues overlaid, which closely matches the eigenvalues of the full latent-state-and-behavior model (last column, top row). But the longest time scales of the full model (bottom, gray arrow) come not from any behavior-only $\vec{H}$, but from slow dynamics associated with transitions between latent states.}
  \label{fig:hierarchy}
\end{figure*}

Thus far we have ignored a major impetus for technical advances: non-Markovianity in observable-behavior-only models \citep{berman_fly_2016}, which can be addressed by fitting a hierarchical model with latent states \citep{weinreb_spontaneous_2026}. The addition of latent states only really changes our definition of `state'---it now means a specific behavior \textit{and} latent state. On the other hand, because latent state transitions occur on a time scale somewhat slower than behavioral transitions, these larger Markov chains are endowed with special structure. We briefly discuss the consequences of this structure below.

\subsection{Whole properties versus part properties}

One striking consequence of combining a latent-state-only chain with a behavior-only chain is that, even if both chains are reversible, the larger chain may not be---and hence, overall behavior can be sequence-like. To understand this point, consider a simple example in which the animal possesses only two possible behaviors and two possible latent states. The latent-state-only and behavior-only chains must satisfy detailed balance, because all two-state chains do. But the overall chain can be cyclic, and hence arbitrarily irreversible. For example, consider the four-state model with transitions
\begin{equation*}
( s, b) \to (s, b') \to (s', b') \to (s', b) \to (s, b) \to \cdots \ .
\end{equation*}
For similar reasons, it is also possible to obtain a reversible model by combining irreversible chains. Hence, qualitatively new properties can appear at the level of the full model. This also means that it is not a priori obvious whether enlarging the state space will increase or decrease effective dimensionality, since highly irreversible models (like the cyclic model) can be low-dimensional.

\subsection{Spectral structure of hierarchical chains}

If latent state and behavioral transitions were completely independent, the spectral structure of the larger chain would be extremely simple: the eigenvalues of the full $\vec{H}$ would be a \textit{sum} of the eigenvalues of the state-only and behavior-only matrices, and its eigenvectors would be a tensor product of each matrix's eigenvectors. Interestingly, something similar is true when latent states and behavior can influence one another, provided that each type of transition occurs on very different time scales: the eigenvalues of the bigger chain are approximately the union of the eigenvalues associated with each latent state's behavior-only chain, and the eigenvectors of the bigger chain correspond to fast behavioral-change modes and slow state-change modes. This result follows from singular perturbation theory \citep{yin2012continuous}, and may not be accurate if the time scale separation is not sufficiently large. This structure appears in the hierarchical model from \citet{weinreb_spontaneous_2026}: the bulk of the spectrum of the full open-field mouse behavior model (Fig. \ref{fig:hierarchy}, last column, top) looks like a union of the spectra associated with each latent state (Fig. \ref{fig:hierarchy}, top row). But the slowest time scales, which are not present in any of the behavior-only models (Fig. \ref{fig:hierarchy}, bottom row), are instead inherited from latent state dynamics.

\section{Discussion}

We explored various consequences of Markov models of behavior, with a particular focus on quantifying its sequence-like character and effective dimensionality. The main contribution of this work is not a new model of behavior, but a clarification of what \textit{must} be true if Markov models are a good description of it.

Several of the consequences we identify do not immediately follow from the observation that the structure of $\vec{H}$ reflects the high-level structure of behavior. For example, our analysis shows that there are multiple coherent ways to define the effective dimensionality of behavior, with one notion reflecting dynamical structure and another relating to predictability. This distinction becomes especially important for highly irreversible models, since these notions can become substantially different. More generally, our results suggest that taking Markov models seriously as descriptions of behavior can help uncover conceptual issues that are easy to overlook when such models are treated purely as descriptive summaries.

\subsection{Objections and limitations}

At this stage, it is worth explicitly discussing possible objections to the approach we have outlined here. First, by treating behavior as the primary object of study, are we making the same mistakes as the behaviorists \citep{skinner_are_1950,chomsky_1959}? The answer is no. That our approach works at all depends on taking one of the major lessons of cognitive science and neuroscience---the idea that behavior ought to be understood in terms of internal states---seriously, since Markov models of behavior \textit{without} latent states tend not to fit \citep{costa_elegans_2024,gautam_zebrafish_2024,weinreb_spontaneous_2026}.

Second, what do we learn from fitting and analyzing Markov models? Why quantify time scales, irreversibility, and dimensionality when quantifying them is not the primary objective? We think there are two reasons. Doing so allows us to make more precise claims; an animal's behavior is not just low-dimensional, but (e.g.) ten times lower-dimensional than when the animal is in some other state. These quantities also furnish what might fairly be called `statistics' of behavior that can be compared across conditions. When the environment changes, we can describe associated behavioral changes not just qualitatively, but in terms of quantitative changes in irreversibility, dimensionality, and so on.

Third, how sensitive are these analyses to the choice of granularity? Most complex syllables can be decomposed into smaller movements (e.g., walking consists of many left-foot and right-foot steps), and many types of movements (e.g., a reach) can be further divided into subtypes (e.g., a reach to the left or right), so it seems unlikely that we will learn anything biologically meaningful if our results depend strongly on how we have chosen to parcellate behavior. In this paper, we have chosen to neglect the question of whether one model of behavior is more appropriate than another, and have concentrated on the simpler question of what a given model implies. But this objection to our approach is significant, because it is almost trivially true that one's results depend strongly on the parcellation of behavior used. For example, when it is possible to coarse-grain behavior so that there are only two states---`active' and `inactive', say---our analyses would conclude that the system is reversible and low-dimensional. Meanwhile, if we treated a very large number of behavioral states as distinct, we would likely come to the opposite conclusion. 

This is a fundamental limitation of our approach: what we learn depends on the assumed model, which includes the details of how behavior is coarse-grained. But this is not necessarily a fatal flaw for two reasons. Principles of parsimony may suggest that a certain coarse-graining balances goodness of fit against model complexity. Also, if multiple levels of description are of interest, one can simply analyze each of the corresponding Markov models and get numbers (e.g., dimensionalities) associated with each scale. This idea is in the spirit of analyzing `barcodes' (collections of some statistic across scales), a strategy used in topological data analysis \citep{chaudhuri_intrinsic_2019,curto_topological_2025}. 

Finally, an important limitation of the analyses we have discussed is that they may become practically unfeasible when the number of states and behaviors becomes too large. This may happen in a few ways: behavior may not have been sufficiently coarse-grained (e.g., too many different reaches are treated as distinct); explaining data may require introducing too many latent states; or the environment is too complicated. Of these, environmental complexity is perhaps the easiest way to make state space prohibitively large. Outside of dark open-field contexts where not much is happening, the dynamics of environmental transitions ought to be included in a Markov description of an animal's behavioral transitions if they significantly modulate behavior. In the worst case, the overall Markov model's size may scale like $\approx (N_E \cdot N_S \cdot N_B)^2$, where $N_E$ is the number of distinct environmental states and $N_S$ is the number of distinct latent states. In addition to making certain analyses (e.g., of eigenvalues and eigenvectors) slower, large increases in the complexity of the environment may make the inference of a Markov model from data extremely difficult. Hence, while Markov models may be indispensable for studying spontaneous behavior in simple environments, the analyses discussed here may be less useful in richer environments.

\subsection{RL and going beyond animal behavior}

At first glance, Markov models appear to provide a slightly different picture of animal behavior than reinforcement learning (RL) \citep{sutton2018reinforcement}. First, in RL behavioral policies are typically conditioned on the animal's current state rather than their current or previous action; hence, accommodating Markov models in an RL framework requires that the notion of state include information about previous actions. This is not unprecedented, as treating recent actions as part of what we mean by `state' can be beneficial in robotics \citep{rado_robo_2024} and for efficiently learning temporally-extended action sequences \citep{bacon_2017}.

Second, the (current) Markov model picture is descriptive rather than normative, in the sense that it does not say why an animal's $\vec{H}$ matrix looks the way it does. One way to connect our picture with RL, especially in the absence of an experimenter-defined task and reward structure, is to assume that the animal acts in a way that maximizes intrinsic rewards \citep{gershman_sub_2025}. For example, if information is intrinsically rewarding \citep{bromberg-martin_neural_2024,bussell_representations_2024}, animals may be driven to investigate novel objects.

Finally, we note that our prescription for studying behavior is somewhat agnostic to the choice of animal, the time scale on which their behavior is monitored, and the extent to which their behavior is abstracted, since we only require that a Markov model provides a good description. This approach can hence be applied not just to model organisms like mice, but to more exotic possibilities, including artificial agents trained to perform tasks like foraging \citep{simmons-edler2025deep}. Furthermore, some of the metrics we have discussed (e.g., entropy production rate) can in principle be computed and fairly compared across wildly different animals and contexts.

Artificial agents are a particularly interesting possible object of study, since although it is easy to access and track their moment-to-moment action sequence (e.g., move one tile left, move one tile right), we may be more interested in emergent behavior better described in terms of high-level actions like `search for food' and `flee from predator'. Such a higher-level description of behavior could plausibly be quantitatively treated via a Markov model, which implies that it may be fruitful to study performant artificial agents just like animals.

\clearpage

\section{Acknowledgments}

KR was funded by the NIH (RF1DA056403, U01NS136507), James S. McDonnell Foundation (220020466), Simons Foundation (Pilot Extension-00003332-02), McKnight Endowment Fund, CIFAR Azrieli Global Scholar Program, NSF (2046583), a Harvard Medical School Neurobiology Lefler Small Grant Award, and a Harvard Medical School Dean’s Innovation Award. This work has been made possible in part by a gift from the Chan Zuckerberg Initiative Foundation to establish the Kempner Institute for the Study of Natural and Artificial Intelligence at Harvard University.

\section{Code availability}

Code for reproducing this paper's figures is available at \url{https://github.com/john-vastola/markov-ccn26}.


\printbibliography

\clearpage

\appendix

\onecolumn

\section{Appendix A. Technical assumptions restricting permissible models}
\label{app:technical}

The set of all possible infinitesimal stochastic matrices $\vec{H}$---that is, all $N \times N$ matrices whose off-diagonal entries are nonnegative and whose columns sum to zero \citep{baez_book_2018}---is quite large, and includes many transition matrices that we do not expect to be good models of behavior. In any case, regardless of their quality, our analysis does not apply to them. We want to exclude
\begin{itemize}
\item models which have states that it is impossible to enter (i.e., behaviors the animal will never do again);
\item models which have states that it is impossible to leave (i.e., behaviors the animal will never stop doing); and
\item models for which the available set of behaviors is initialization-dependent.
\end{itemize}
If we do not exclude such models, it is trivially possible to have multiple steady state distributions, or for the steady state distribution to contain zeros or ones. To avoid these possibilities, we will assume that $\vec{H}$ is \textit{irreducible}. In terms of the graph associated with $\vec{H}$, irreducibility is equivalent to demanding that any node is reachable from any other node via some directed path through the graph. Put differently, there are not multiple isolated `islands' of behavior, and there are no source or sink behaviors.

If $\vec{H}$ is irreducible, the Perron-Frobenius theorem applies, and tells us that there exists a unique eigenvector of $\vec{H}$ that (i) has eigenvalue zero, (ii) has nonnegative entries, and (iii) sums to one. Moreover, this vector (which we denote by $\vec{\pi}$) is the steady state distribution, and it is always true that $\lim_{t \to \infty} \vec{p}(t) = \vec{\pi}$ regardless of initial condition. We also assume that $\vec{H}$ is diagonalizable (which is not necessarily true for a generic irreducible infinitesimal stochastic matrix), but this is only to guarantee that we can sensibly talk about eigenvalues and eigenvectors.

Readers interested in a deeper mathematical understanding of these restrictions, and of Markov chains more generally, should consult some combination of \citet{hairer2006ergodic}, \citet{baez_book_2018}, and \citet{seabrook_tutorial_2023}. Our continuous-time formulation most closely follows that of \citet{baez_book_2018}.

\clearpage

\section{Appendix B. Toy models of behavior}
\label{app:toymodels}

In this appendix, we define and compute basic properties of several toy models of behavior.

\subsection{The bag model}

The \textbf{bag model} assumes all possible behaviors are equally likely, and has a transition matrix
\begin{equation}
\vec{H} := \frac{k}{N} \begin{pmatrix}
- (N - 1) & 1 & \cdots & 1 \\
1 & - (N - 1) & \cdots & 1 \\
\vdots & & & \\
1 &  1 & \cdots &- (N - 1)
\end{pmatrix} = - k \ \vec{I} + \frac{k}{N} \vec{1} \vec{1}^T \ .
\end{equation}
We can compute its characteristic polynomial (and hence its eigenvalues) using the matrix determinant lemma:
\begin{equation}
\begin{split}
\det\left[  - (k + \lambda) \vec{I} + \frac{k}{N} \vec{1} \vec{1}^T \right] &= [- (k + \lambda) ]^N \left[  1 - \frac{1}{N} \vec{1}^T \vec{1} \frac{k}{(k + \lambda)}  \right] = (-1)^N  (k + \lambda)^{N-1} \lambda   \ .
\end{split}
\end{equation}
Hence, there are $N-1$ eigenvalues with $\lambda = - k$ and one with $\lambda = 0$. Any vector orthogonal to $\vec{1}$ is an eigenvector with eigenvalue $\lambda = - k$; the dimensionality of this space of vectors is $N-1$. 

The steady state distribution for this model is uniform, since $\vec{1}$ is an eigenvector with eigenvalue zero.

\subsection{The cyclic model}

\noindent Define the $N \times N$ \textbf{shift matrix} $\vec{S}$ via
\begin{equation}
S_{i j} = \begin{cases}
1 & \text{ if } i = j + 1 \\
1 & \text{ if } i = 1 \text{ and } j = N \\
0 & \text{ otherwise}
\end{cases} \ .
\end{equation}
This matrix represents a cyclic shift of any vector, since
\begin{equation}
\vec{S} \begin{pmatrix}
v_1 \\ v_2 \\ \vdots \\ v_{N-1} \\ v_N \end{pmatrix} = \begin{pmatrix}
v_N \\ v_1 \\ \vdots \\ v_{N-2} \\ v_{N-1} \end{pmatrix}  \ .
\end{equation}
The transpose of $\vec{S}$ represents a cyclic shift in the other direction:
\begin{equation}
\vec{S}^{-1} \begin{pmatrix}
v_1 \\ v_2 \\ \vdots \\ v_{N-1} \\ v_N \end{pmatrix} = \begin{pmatrix}
v_2 \\ v_3 \\ \vdots \\ v_{N} \\ v_{1} \end{pmatrix}  \ .
\end{equation}
This implies that $\vec{S}^T = \vec{S}^{-1}$. Define the \textbf{cyclic model} via the $N \times N$ transition matrix
\begin{equation}
\vec{H} = - (k_f + k_r) \vec{I} + k_f \vec{S} + k_r \vec{S}^{-1} \ ,
\end{equation}
where the forward rate $k_f$ and reverse rate $k_r$ are nonnegative constants, at least one of which is nonzero. This model abstracts the idea of transitions along a `ring' of $N$ states. If $k_f$ and $k_r$ are the same, then transitions are `diffusive' in the sense that transitions do not tend to have any directionality. Meanwhile, if one of the rates is much larger than the other, transitions tend to take place in a fixed order. For example, if $k_f$ is the significantly larger one, states tend to be visited in the $1 \to 2 \to ... \to N$ order.

\paragraph{Eigenvectors and eigenvalues of the shift matrix.} It is well known that the eigenvectors of the shift matrix correspond to Fourier modes. To see this, let $\omega := e^{\frac{2 \pi i}{N}}$, and define the vector $\vec{v}_k$ (for $k = 0, ..., N-1$) via
\begin{equation}
v_{k \ell} := \frac{1}{\sqrt{N}} \omega^{-k (\ell - 1)} 
\end{equation}
for all $\ell = 1, ..., N$. Since 
\begin{equation}
\vec{S} \vec{v}_k = \omega^{-k (N-1)} \vec{v}_k = (\omega^{-1})^{-k} \vec{v}_k = \omega^k \vec{v}_k \ ,
\end{equation}
$\vec{v}_k$ is an eigenvector of $\vec{S}$ with eigenvalue $\omega^k$. Since we have precisely $N$ such vectors, these are all of the eigenvectors and eigenvalues of $\vec{S}$. Relatedly, observe that
\begin{equation}
\vec{S}^T \vec{v}_k = \omega^{-k} \vec{v}_k \ ,
\end{equation}
so the eigenvalues of $\vec{S}^T$ are the `opposite' of those of $\vec{S}$ in their assignments.

\paragraph{Eigenvectors and eigenvalues of the transition matrix.} By definition, $\vec{H}$ is a linear combination of $\vec{I}$, $\vec{S}$, and $\vec{S}^{-1}$, which we just learned share eigenvectors. Hence, the $\vec{v}_k$ are also eigenvectors of $\vec{H}$. Its eigenvalues are different: 
\begin{equation}
\vec{H} \vec{v}_k = \left[ - (k_f + k_r) + k_f \omega^k + k_r \omega^{-k}  \right] \vec{v}_k \ .
\end{equation}
Simplifying them, we find
\begin{equation}
\lambda_k = - (k_f + k_r)  + k_f \omega^k + k_r \omega^{-k} = - (k_f + k_r) \left[ 1 - \cos\left( \frac{2 \pi k}{N} \right)  \right] + (k_f - k_r) \sin\left( \frac{2 \pi k}{N}  \right) \ i 
\end{equation}
for $k = 0, 1, ..., N-1$. Note that when $k_f = k_r$, in which case movement along the ring is `diffusive' rather than directed, the imaginary part of these eigenvalues becomes zero. On the other hand, as $|k_f - k_r|$ increases, the imaginary part grows increasingly large.

\subsection{The two-state model}

The \textbf{two-state model} is perhaps the simplest possible non-degenerate Markov chain. As its name implies, it has only two states and some rate of transitions between them. Define its transition matrix via
\begin{equation}
\vec{H} := \begin{pmatrix}
- k_{21} & k_{12} \\
k_{21} & - k_{12}
\end{pmatrix} \ .
\end{equation}

\paragraph{Steady state distribution.} The steady state distribution is
\begin{equation}
\vec{\pi} = \begin{pmatrix}
\frac{k_{12}}{k_{12} + k_{21}} \\
\frac{k_{21}}{k_{12} + k_{21}} 
\end{pmatrix} \ .
\end{equation}

\paragraph{Eigenvalues and eigenvectors.} A trivial calculation tells us that the nonzero eigenvalue of this model is $\lambda = - (k_{12} + k_{21})$, and that the corresponding eigenvector is
\begin{equation}
\vec{v}_1 = \begin{pmatrix} 1 \\ -1 \end{pmatrix} \ .
\end{equation}
This eigenvector is particularly easy to interpret: it represents a change in behavior involving doing behavior 1 more often and behavior 2 less often. Since eigenvectors are only defined up to an overall scaling, it also represents doing behavior 1 less often and behavior 2 more often. This is somewhat trivial, since no other changes in behavior are possible for a model with only two states.

\subsection{The three-state sleep-wake model}

In the main text, we referenced a simple model of sleep with three states, which we will label as `asleep', `awake', and `awake and alert'. Transitions between `asleep' and `awake' tend to happen slowly, and transitions between `awake' and `awake and alert' tend to happen quickly; direct transitions between `asleep' and `awake and alert' are prohibited.

This model is interesting because it provides a simple and interpretable example of a model of behavior that can possess well-separated time scales. Define its transition matrix via
\begin{equation}
\vec{H} = \begin{pmatrix}
- 2 \epsilon & \gamma   &  0 \\
2 \epsilon & - (\epsilon + \gamma + k)   & 2\gamma   \\
0  &  \epsilon + k  & -  2 \gamma
\end{pmatrix}
\end{equation}
where we assume $k \gg \gamma$ (i.e., $k$ is very fast) and $\gamma \gg \epsilon$ (i.e., $\epsilon$ is very slow). Since \citet{vastola_storing_2025} studied this model in detail, we will state their results rather than derive them from scratch. 

\paragraph{Steady state distribution.} The steady state distribution is
\begin{equation}
 \vec{\pi} = \frac{1}{(\epsilon + \gamma)^2 + k \epsilon} \begin{pmatrix} \gamma^2 \\ 2 \gamma \epsilon \\ \epsilon^2 + k \epsilon \end{pmatrix}  \ .
\end{equation}

\paragraph{Eigenvalues and eigenvectors.} The eigenvalues and eigenvectors of this model depend on the `spectral gap'
\begin{equation}
\Delta := \frac{\sqrt{(\epsilon + \gamma + k)^2 + 4 k (\gamma - \epsilon)} - (\epsilon + \gamma + k)}{2} \geq 0 \ .
\end{equation}
In terms of this quantity, the nonzero eigenvalues of $\vec{H}$ are
\begin{equation}
\lambda =  - (\epsilon + \gamma) + \Delta, - 2 (\epsilon + \gamma) - \Delta - k \ .
\end{equation}
In the large $k$ limit, these correspond to well-separated time scales, since (using $\Delta \approx \gamma - \epsilon$)
\begin{equation}
\lambda \approx - 2 \epsilon, - (3 \gamma + \epsilon + k) \ .
\end{equation}
In particular, $\tau_1 \approx 1/(2 \epsilon)$ is very long, and $\tau_2 \approx 1/k$ is very short. The corresponding eigenvectors are
\begin{equation}
\vec{v}_1 = \frac{1}{\gamma} \begin{pmatrix}
\gamma \\ - \gamma + \epsilon + \Delta  \\ - \epsilon - \Delta
\end{pmatrix} \hspace{0.5in} \vec{v}_2 =  \begin{pmatrix}
\frac{1}{2} \left[ 1 - \frac{\Delta}{\gamma - \epsilon} \right] \\ - 1  \\ \frac{1}{2} \left[ 1 + \frac{\Delta}{\gamma - \epsilon} \right]
\end{pmatrix}  \ .
\end{equation}
These are somewhat interpretable, especially in the large $k$ limit. For very large $k$, 
\begin{equation}
\vec{v}_1 \approx  \begin{pmatrix}
1 \\ 0 \\ - 1
\end{pmatrix} \hspace{0.5in} \vec{v}_2 \approx  \begin{pmatrix}
0 \\ - 1  \\ 1
\end{pmatrix}  \ .
\end{equation}
Hence, the fast mode $\vec{v}_2$ moves probability between the `awake' and `awake and alert' states only, while the slower $\vec{v}_1$ mode moves probability between the `asleep' and `awake and alert' states.

\clearpage

\section{Appendix C. Hierarchical model fit to real data}

\citet{weinreb_spontaneous_2026} fit a hierarchical hidden Markov model (using an approach they called \textit{shMoSeq}) to open-field spontaneous mouse behavior, and recovered a transition matrix $\vec{P}^T$ that specifies the small-time probability of transitioning from one syllable and latent state to another. Their `mesoscopic' time scale $\Delta t$ was effectively $1/30$ seconds, since behavior was recorded at 30 frames per second.

To convert their $\vec{P}^T$ into an infinitesimal stochastic $\vec{H}$, we moved from their row-stochastic convention (i.e., rows sum to one) to a column-stochastic convention by transposing their matrix, and then exploiting the relationship
\begin{equation}
\vec{P} = e^{\vec{H} \Delta t} \approx \vec{I} + \vec{H} \Delta t \ ,
\end{equation}
which becomes increasingly accurate as $\Delta t$ gets smaller. We did this instead of deriving $\vec{H}$ via a matrix logarithm, partly because we empirically found that doing so led to numerical instability.

We used \textit{only} their model labeled `solitary\_and\_social\_open\_field', and not any of the other available models. This is partly because this model was simplest, since it involves only a small number of latent states, all of which have been annotated. This model has $100$ behavioral syllables and $5$ latent states. The latent states were assigned the labels `grooming', `exploratory locomotion 1', `exploratory locomotion 2', `local investigation', and `social engagement', and we have retained those labels here.

\clearpage

\section{Appendix D. Motivation for predictive dimensionality}

Here, we derive the approximation to mutual information exploited in the main text to define the predictive dimensionality of a model. Let $\vec{P}(\tau) := e^{\vec{H} \tau}$ denote the matrix controlling the probability of transitioning from one state to another after an amount of time $\tau$. The mutual information between the system's present state (assuming it is stationary) and the system's state after an amount of time $\tau$ has passed is defined as
\begin{equation}
I(\tau) := \sum_{a, b} P_{b a} \pi_a \log\left( \frac{P_{ba} \pi_a}{(\sum_c P_{b c} \pi_c   ) \pi_a }  \right) = \sum_a \pi_a \sum_{b} P_{b a}  \log\left( \frac{P_{ba}}{(\sum_c P_{b c} \pi_c   ) }  \right) = \sum_a \pi_a \sum_{b} P_{b a}  \log\left( \frac{P_{ba}}{\pi_b}  \right)  \ ,
\end{equation}
where we used the fact that $\vec{\pi}$ is the steady state distribution in the last step to write $\pi_b = \sum_c P_{b c} \pi_c$. We will approximate this quantity by assuming $P_{ba}$ is not far from steady state, so that $\delta p_{b a} := P_{b a} - \pi_b$ is small. (More precisely, we assume $\delta p_{b a}/\pi_b \ll 1$.) This assumption is useful because it allows us to Taylor expand $I(\tau)$ to second order in $\delta p_{b a}$. Doing so, we find
\begin{equation}
\begin{split}
I(\tau) &= \sum_a \pi_a \sum_{b} \pi_b \left(  1 + \frac{\delta p_{b a}}{\pi_b} \right)  \log\left(  1 + \frac{\delta p_{b a}}{\pi_b}  \right)  \\
&\approx \sum_a \pi_a \sum_{b} \pi_b \left(  1 + \frac{\delta p_{b a}}{\pi_b} \right) \left( \frac{\delta p_{b a}}{\pi_b} - \frac{1}{2} \frac{\delta p_{b a}^2}{\pi_b^2} \right) \\
&\approx \sum_a \pi_a \sum_{b} \pi_b  \left( \frac{\delta p_{b a}}{\pi_b} + \frac{1}{2} \frac{\delta p_{b a}^2}{\pi_b^2} \right)  \ .
\end{split}
\end{equation}
Since $\sum_b \delta p_{b a} = \sum_b (P_{b a} - \pi_b) = 0$, this simplifies to
\begin{equation}
\begin{split}
I(\tau) &\approx \frac{1}{2}  \sum_{a, b}  \delta p_{b a}^2 \frac{\pi_a}{\pi_b}  =  \frac{1}{2}  \sum_{a, b} ( P_{b a} - \pi_b )^2 \frac{\pi_a}{\pi_b}=  \frac{1}{2}  \sum_{a, b}  P_{ba}^2 \frac{\pi_a}{\pi_b}  - \frac{1}{2} \ .
\end{split}
\end{equation}
But this means
\begin{equation}
I(\tau) \approx \frac{1}{2} \text{tr}\left[ \vec{D}^{1/2} \vec{P}^T \vec{D}^{-1} \vec{P} \vec{D}^{1/2}  \right] - \frac{1}{2} = \frac{1}{2} \text{tr} \vec{K}(\tau) - \frac{1}{2} \ ,
\end{equation}
where we define the \textit{past-future kernel} via
\begin{equation}
\vec{K}(\tau) := \vec{D}^{1/2} e^{\vec{H}^T \tau} \vec{D}^{-1} e^{\vec{H} \tau} \vec{D}^{1/2}   \ .
\end{equation}
$\vec{K}(\tau)$ is clearly real, symmetric, and positive definite; this implies that its eigenvalues $\nu_k$ are real and positive. Moreover, the Perron-Frobenius theorem implies that $1$ is always an eigenvalue, and that all other eigenvalues are less than $1$. Since $1$ is an eigenvalue, we can use the $\nu_k = 1$ term to cancel the offset, so that
\begin{equation}
I(\tau) \approx \frac{1}{2} \sum_{\nu_k < 1} \nu_k \ .
\end{equation}
This expression for $I(\tau)$ is a sum of positive eigenvalues, each of which we can interpret as `explaining' part of the overall mutual information between the present and future.

\end{document}